\documentclass[useAMS,usenatbib,twocolumn]{mnras}
\usepackage{natbib}
\usepackage{graphicx}
\usepackage{times}
\usepackage{rotate}
\usepackage{hyperref}
\usepackage{url} 

\newcommand{\asec}{\arcsec}

\newcommand{\ds}{\ensuremath{\Delta\Sigma}}

\newcommand{\alexie}[1]{\textbf{\color{alexie} }}

\begin{document}
  

\title[Dwarf Lensing]{Deep$+$Wide Lensing Surveys will Provide Exquisite Measurements of the Dark Matter Halos of Dwarf Galaxies}





\author[Leauthaud et al.]  {Alexie Leauthaud$^{1}$, Sukhdeep Singh$^{2}$, Yifei Luo$^{1}$, Felipe Ardila$^{1}$, Johnny~P.~Greco$^{3}$,\newauthor Peter Capak$^{4,5}$, Jenny~E.~Greene$^6$,  Lucio Mayer$^{7}$
\\
$^1$Department of Astronomy and Astrophysics, University of California, Santa Cruz, 1156 High Street, Santa Cruz, CA 95064 USA
\\
$^2$Berkeley Center for Cosmological Physics, Department of Physics, University of California, Berkeley \& Lawrence Berkeley National Laboratory,\\ CA 94720, USA
\\
$^3$Center for Cosmology and AstroParticle Physics (CCAPP), The Ohio State University, Columbus, OH 43210, USA
\\
$^4$IPAC, California Institute of Technology, 1200 East California Blvd., Pasadena, CA 91125, USA
\\
$^5$Cosmic Dawn Center (DAWN)
\\
$^6$Department of Astrophysical Sciences, 4 Ivy Lane, Princeton University, Princeton, NJ 08544
\\
$^7$CTAC, Institute for Computational Science, University of Zurich, Winterthurerstrasse 190, 8057 Zurich, Switzerland}

\maketitle
\label{firstpage}

 
\begin{abstract} The advent of new deep$+$wide photometric lensing surveys will open up the possibility of direct measurements of the dark matter halos of dwarf galaxies. The HSC wide survey will be the first with the statistical capability of measuring the lensing signal with high signal-to-noise
at $\log(M^*)\sim 8$. At this same mass scale, LSST will have the most overall constraining power with a predicted signal-to-noise for the galaxy-galaxy lensing signal around dwarfs of S\/N$\sim$200. WFIRST and LSST will have the greatest potential to push
below the $\log(M^*)=7$ mass scale thanks to the depth of their imaging data. Studies of the dark matter halos of dwarf galaxies at $z\sim$0.1 with gravitational lensing are soon within reach. However, further work will be required to develop optimized strategies for extracting dwarfs samples from these surveys, determining redshifts, and accurately measuring lensing on small radial scales. Dwarf lensing will be a new and powerful tool to constrain the halo masses and inner density slopes of dwarf galaxies and to distinguish between baryonic feedback and modified dark matter scenarios.
\end{abstract}

 

 
 



\section{Introduction}

Dwarf galaxies are a unique probe of the nature of dark matter and of the interplay between dark matter and baryonic physics. The long standing cusp-core controversy, whereby the rotation curves of many gas-rich dwarfs (dwarf spirals and dwarf irregulars) favor flatter dark matter profiles relative to the cuspy profiles predicted by Cold Dark Matter (CDM), can be explained by several competing scenarios \citep[e.g.,][]{Pontzen:2014aa}. Some models invoke modifications of CDM, such as self-interacting dark matter (SIDM), while other models rely on baryonic physics, such as supernovae-driven outflows which can modify the inner slope of the dark matter profile via potential fluctuations \citep[][]{Governato:2010aa, Pontzen:2012aa, Onorbe:2015aa,Di-Cintio:2014aa, Tollet:2016aa}. All of these competing models predict a flattening of the innermost dark matter density profile (higher values of $\alpha$ where $\rho_{\rm DM} \propto r^\alpha$) on scales of 0.5-1 kpc (referred to as the ``core'' region). These models yield a better description of the observed kinematics of dwarf galaxies \citep[e.g.,][]{Oh:2015aa} which favor $\alpha\sim -0.3$ over the CDM prediction of $\alpha=-1$. A flattening of the inner halo profile may also solve other long-standing issues of CDM-based structure formation at small scales, such as the too-big-to-fail problem \citep[e.g.,][]{Brooks:2013aa}, perhaps in combination with tidal effects \citep[][]{Tomozeiu:2016aa}. 

By design, all of the models predict a flattening of the inner dark matter profile for dwarfs. Hence, measurements of $\alpha$ alone are insufficient to distinguish between such models; additional observables are required. Baryonic feedback models predict a strong connection between the flattening of the inner dark matter slope, $\alpha$, and galaxy properties \citep[stellar mass and star formation efficiency or burstiness of the star formation rate,][]{Governato:2010aa, Onorbe:2015aa,Di-Cintio:2014aa, Tollet:2016aa}. Non baryonic solutions to the cusp-core controversy on the other hand, do not predict such correlations. A detection of these correlations would therefore be a powerful argument in favor of  baryonic feedback models over modifications to CDM. Recent theoretical work has shown that supernovae-driven outflows have the strongest impact on the inner dark matter slope $\gamma$ in the mass range $10^8-10^{10}$ $M_{\odot}$. At lower mass scales,  star formation is too inefficient to generate significant mass displacement via outflows while at larger mass scales, the
potential well is too deep for outflows to be effective at generating potential fluctuations \citep[][]{Governato:2012aa,Chan:2015aa,Tollet:2016aa}. Hence, the mass range $M_*=10^8-10^{10}$ $M_{\odot}$ is a ``sweet spot'' in terms of trying to detect correlations between dwarf properties, the inner dark matter halo slope, and dark matter halo mass.

Galaxies properties are straightforward to measure, and kinematic studies can be used to probe the inner slope $\alpha$.  But the total halo mass is the key missing component required to complete this picture. The THINGS and LITTLE-THINGS 21cm HI surveys, which focus on galaxies in the range $M_*=10^8-10^{10} M_{\odot}$, only measure the rotation curves of dwarfs on scales up to $R_{\rm max}=5-10$ kpc \citep[][]{Oh:2015aa}. Hence rotation curves of dwarf galaxies only yield a measurement of the total mass on scales below  $\sim$ 10-20 kpc. This is a factor of $\sim$10-20 smaller than the actual halo radius ($R_{\rm 200b}\sim 90-150$ kpc at $M_*\sim 10^{8.5}$ M$_{\odot}$). Any ``halo mass'' estimate from rotation curves is in fact an extrapolation that relies on assumptions about the shape of the dark matter profile \citep[e.g.,][]{Kazantzidis:2004aa}. Furthermore, cosmological hydro simulations of dwarfs suggest that the inner profiles of dwarfs display a wide range of slopes with values ranging from $\alpha=-1$ to $\alpha=-0.3$ \citep[][]{Chan:2015aa,Tollet:2016aa}. This would imply that conventional measurements of kinematics measurements simply cannot be used to determine halo masses (a large dispersion in $\alpha$ would mean that one cannot extrapolate to larger scales because there is not a single universal halo profile). Similar issues also apply to stellar kinematics studies of gas poor dwarf spheroidals and dwarf ellipticals.  In short, the scales on which both rotation curves and stellar kinematics can be measured only provide insight on the inner dark matter profile.

For the reasons outlined above, independent and large scale measurements of the dark matter profile would be tremendously powerful and highly complementary to small-scale kinematic studies of dwarfs. The combination of a large scale halo mass estimate, together with rotation curve data, or stellar kinematic data, would yield much more accurate constraints on both the total halo masses $M_{\rm halo}$ of dwarfs as well as their inner dark matter slopes $\alpha$. The lack of total halo mass measurements is the key missing ingredient that is required in order to full understand the interplay between baryonic physics and dark matter profiles in dwarf galaxies. 

One of the most powerful ways to directly probe total halo masses out to the halo radius is via weak gravitational lensing.  In particular, the ``galaxy-galaxy lensing'' technique measures the average weak lensing signal from background ``source'' galaxies around a sample of foreground ``lens'' galaxies (typically several hundred to thousands of lens galaxies). Galaxy-galaxy lensing is one of the most effective techniques that can be used to  measure the full dark matter profile of galaxies, from scales of a few tens of kpc to scales of several Mpc. However, existing weak lensing measurements have been limited to galaxies with $M_*>10^{9}$ $M_{\odot}$ \citep[e.g.,][]{Leauthaud:2012}.

The goal of this paper is to demonstrate that the advent of lensing surveys that are both \emph{deep and wide} will enable the discovery of large enough samples of $z\sim 0.1$ dwarfs for direct measurements of the dark matter halos of dwarf galaxies with galaxy-galaxy lensing. We present forecasts for the signal-to-noise of galaxy-galaxy lensing around dwarf galaxies for the Hyper Suprime Cam survey \citep[HSC,][]{Aihara:2018}, and for upcoming surveys such as the Large Synoptic Survey Telescope \citep[LSST,][]{Ivezic:2008}, \emph{Euclid} \citep{Laureijs:2011}, and the Wide Field Infrared Survey Telescope \citep[WFIRST,][]{Spergel:2013}. Section \ref{section:comp} presents an estimate of the mass completeness limits of these surveys. Section \ref{forecasts} presents our methodology and Section \ref{results} presents forecasts. Section \ref{conclusions} presents a summary and our conclusions. We assume a cosmology with $\Omega_\lambda=0.693$, $\sigma_8=0.823$, $H_0=67.8$. We use physical units for the lensing observable $\Delta\Sigma$.

\section{Completeness Limits of Upcoming Surveys}\label{section:comp}

Deep$+$wide photometric data will be required to identify sufficient numbers of dwarf galaxies to measure halo masses with lensing. In this paper, we assume that the dwarf lens samples will be selected from the same imaging data used for shape measurements. For this reason, we begin by estimating the stellar mass completeness limits of lensing surveys. Here, we consider five lensing surveys:  COSMOS, HSC, LSST, \emph{Euclid}, and WFIRST. We begin by considering existing surveys (COSMOS and HSC). We then use these results to estimate the completeness limits for future surveys. 

\subsection{COSMOS, HSC, and Surface Brightness Effects}

The COSMOS survey \citep{Scoville:2007} provides more than 30 bands of deep imaging data, spanning UV to radio wavelengths. The COSMOS2015 catalog presents the latest public data release for the COSMOS survey \citet{Laigle:2016aa}. The COSMOS $i$-band 5 sigma point source depth is $i=25.9$ (C. Laigle, priv. comm). 

The HSC survey is an ongoing effort that aims to map 1400 deg to $i\sim$26 in $grizy$ using the 8.2m Subaru telescope. The depth of the HSC wide survey is $i=26.4$ (5$\sigma$ point source) \citep{Aihara:2018}. The COSMOS $i$-band data is slightly shallower than HSC wide, but for simplicity, we will assume for the remainder of this paper that the HSC and COSMOS have roughly the same mass sensitivity at $z<0.3$.
 
The completeness of the COSMOS survey  has already been well characterized for mass function studies. We use the mass completeness limits estimates from \citet{Laigle:2016aa} (hereafter L16) who performed a detailed analysis of the completeness of COSMOS in order to measure the galaxy mass function. In brief, they first estimate the photometric errors for each of their bands by placing apertures on empty portions of the sky in 2 and 3\asec apertures and measuring the noise distribution.  Second, a model grid of SEDs was compared with the $K$-band limit to determine the 90\% completeness limit for each stellar mass.  Finally, they derive a functional form for the completeness as a function of redshift scaled to the depth of the $Ks$-band data.  The estimated COSMOS completeness limits are shown in Figure \ref{comp} which is for an estimated $K$-band depth of 25.0 $5\sigma$ in a 2\asec aperture. COSMOS is mass complete to $\log(M_*)= \sim 7.3$ at $z=0.1$ and to $\log(M_*)= \sim 8.1$ at $z=0.3$.

Since the COSMOS completeness estimates from L16 are derived from a fixed aperture, the effect of surface brightness sensitivity is not explicitly included. At fixed stellar mass, dwarf galaxies are observed to span a wide range of sizes \citep[e.g.,][]{McConnachie:2012}, leading to a range of surface brightness values, which will impact the mass completeness \citep[e.g.,][]{Blanton:2005aa}. To investigate the importance of this effect, we use the pipeline\footnote{\url{https://github.com/johnnygreco/hugs}} from \citet{Greco:2018} to inject PSF-convolved Sersic functions with a range of sizes ($r_\mathrm{reff} = 2\asec$-$10\asec$) and surface brightnesses (22-29~mag arcsec$^{-2}$) into HSC survey images across the entire footprint of the survey. This pipeline was designed specifically to detect extended low surface brightness galaxies in the HSC survey. We assign each mock galaxy a stellar mass by sampling the stellar mass--surface brightness relation (including scatter) from \citet{Danieli:2018}, which is derived from dwarf galaxies in and around the Local Group. We recover the mock galaxies with a 90\% mass limit of $\log M_*/M_\odot \sim 7.3$ at $z=0$, with relatively little dependence on galaxy effective radius. To model the redshift dependence of the completeness, we use the surface brightness completeness function to scale the $z=0$ mass limit according to cosmological surface brightness dimming. The results are indicated by the upper black line in Figure~\ref{comp}.

Given surface brightness effects, we estimate the mass completeness curves for COSMOS/HSC at low-$z$ to be roughly located within the grey shaded region in Figure ~\ref{comp}. At low redshifts, the completeness will depend on both mass, surface brightness, but also on the exact pipeline used to detect objects. Traditional pipelines, such as those used in the SDSS and HSC surveys, are generally excellent for detecting high surface brightness galaxies in non-crowded fields, but they have not been optimized for diffuse dwarf galaxy detection, making them susceptible to surface brightness selection effects  \citep[e.g.,][]{Kniazev:2004}. Further work will be required to optimize detection methods for dwarfs and to characterize more precise completeness limits. We cannot say whether COSMOS or HSC can 
reliably detect dwarfs with $\log M_*/M_\odot \sim 7.0$ at $z\sim 0.05$, however, based on our tests, the detection of dwarfs with $\log M_*/M_\odot > 8.0$ at $z< 0.2$ should be robust.
 
\subsection{LSST, \emph{Euclid}, and WFIRST}

We do not have the same galaxy-injection tools in place yet for other surveys. Hence, to estimate the completeness limits for other surveys, we adopt the following simple approximation. Due to the physics of stellar evolution, the stellar mass is strongly correlated with the rest-frame optical flux in the 0.4-2$\mu$m range (see e.g. \citealt{Bruzual:2003} or \citealt{Maraston:2005}) with a secondary dependence on the age of the stellar population.  So the primary survey characteristic of interest for mass completeness is the depth of the survey data in this rest frame wavelength range.  At the redshifts of $z<0.3$ we are interested in for this paper the depth in the $i$ band (observed $\sim0.75\mu$m) or the deepest band red-ward of $i$ is a good proxy.  For simplicity we will scale the survey completeness to those depths. As long as the relative depths of the 0.3-1$\mu$m photometry are similar to COSMOS this should be a good proxy at $z<0.3$. We therefore scale the COSMOS completeness limits as $\Delta i/2.5$  where $\Delta i$ it the difference in i-band depth compared to COSMOS (5$\sigma$ point source).

 The LSST \citep[][]{Ivezic:2008} will be a large wide-field ground-based system with a 8.4 m (6.5 m effective) primary mirror. LSST begins operations in 2023 and plans to map out 18,000 deg$^2$. The 10-year 5$\sigma$ point source depth of $i=26.8$ \citep[][]{LSST-Science-Collaboration:2009}. By scaling the $i$-band sensitivity with respect to COSMOS, we find that the  10-year LSST optical imaging will therefore be 0.36 dex more sensitive in mass than COSMOS. 
  
\emph{Euclid} is a European space mission with a 1.2m primary mirror and with an expected launch in 2020 \citep{Laureijs:2011}. Over 6 years,  \emph{Euclid} will conduct both an imaging and a spectroscopic survey over the lowest background 15000 deg$^2$ of the extragalactic sky. The \emph{Euclid} catalog will be selected in a broad $r+i+z$ filter similar to, but wider than, the HST F814W filter. For \emph{Euclid}, the depth in the wide field will be 26.3 ABmag ($5\sigma$ point source, H. Hoekstra et al. priv comm) over 15,000 square degrees. Complimentary data will be obtained in $grizYJH$ bands from the ground and from the \emph{Euclid} Near-Infrared channel. Here we consider the depth of the Euclid $r+i+z$ wide field imaging. For simplicity, we assume here that \emph{Euclid} will have roughly the same mass sensitivity as COSMOS ($i=25.9$) and HSC ($i=26.4$).

\begin{figure}
\centering
\includegraphics[width=8.7cm]{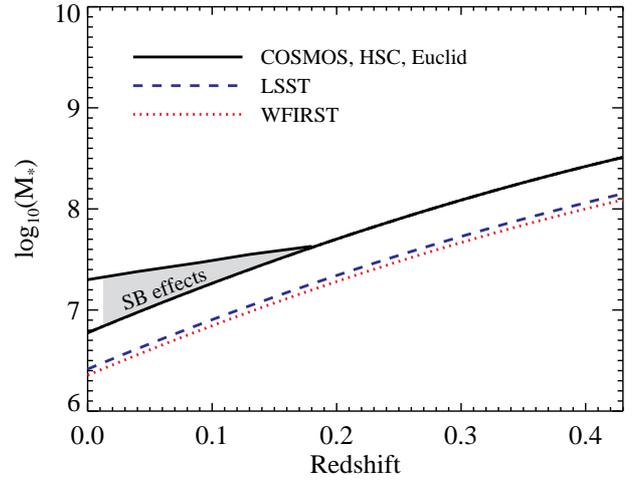}
\caption{Stellar mass completeness of lensing surveys. The black solid line corresponds to the COSMOS2015 catalog. The grey shaded region indicates where surface brightness effects may impact completeness estimates. HSC and \emph{Euclid} have roughly the same mass sensitivity as COSMOS2015. LSST will be more complete than COSMOS2015 by 0.36 dex. WFIRST adds an extra $\approx$0.43 dex in terms of completeness compared to COSMOS2015. Surface brightness effects will need to be investigated in further detail, especially for LSST and WFIRST.}
\label{comp}
\end{figure}

 The \textit{Wide Field Infrared Survey Telescope} (WFIRST) is a 2.4m telescope NASA mission with a launch in 2024 \citep{Spergel:2015}. WFIRST will be NIR selected in the 1-2$\mu$m wavelength range and is anticipated to reach a depth of 25.8-26.7 ABmag over $\sim2,200$ square degrees depending on the filter\footnote{The quoted depth is deeper than the the one typically quoted for lensing because the lensing source catalog typically cuts at S\/N$>20$ which is 1.5 mag brighter than the limits quoted here.}. To estimate the mass completeness of WFIRST, we scale the $i$-band depth as described previously, and add 0.1 dex for the red selection. With this calculation, WFIRST is 0.42 dex more sensitive in mass than COSMOS. 
 
Figure \ref{comp} displays the mass completeness limits of these surveys as a function of redshift. Columns 2 and 3 in \ref{forecasttable} indicate the mass completeness limits for the surveys under consideration at $z=0.1$ and $z=0.3$. Of the surveys under consideration, WFIRST and LSST will have the greatest potential for pushing to low halo mass. They may be capable of detecting dwarf lens galaxies with masses below $\log(M_*)=7$ at $z<0.1$. But further work will be required to investigate the impact of surface brightness effects and to develop adapted detection algorithms.

\begin{table*}
  \caption{Estimated completeness limits and lensing parameters. The completeness limits may be optimistic given possible surface brightness effects. This will be investigated in future work.}
\begin{tabular}{@{}lccccc}
\hline
Survey & $M_*$ limit at & $M_*$ limit at & Area in  & N source & Mean redshift\\
 & $z=0.1$ & $z=0.3$ & deg$^2$ & per arcmin$^2$ & of sources $\langle z_s \rangle$\\
\hline
COSMOS & 7.3  & 8.1  & 1.64 & 39 & 1.2 \\
HSC Wide & 7.3 & 8.1 & 1000 & 18.5 & 0.81 \\
LSST Wide & 6.94 & 7.74 & 18,000 & 30 & 1.2  \\
\emph{Euclid} Wide & 7.3 & 8.1  & 15,000  & 30  & 0.9 \\
WFIRST HLS & 6.9 & 7.68 & 2,400 & 45 & 1.1 \\
\hline
\end{tabular}
\label{forecasttable}
\end{table*}

\section{Forecast Methodology}\label{forecasts}

We now consider how well ongoing and future surveys will be able to measure the galaxy-galaxy lensing signal for dwarf lenses. Here, we set aside the question of how to determine redshifts for dwarfs, as well as the impact of lensing systematic errors. We focus only on estimating the statistical uncertainties on the lensing signal given a dwarf lens sample with known redshifts.

\subsection{Amplitude of Dwarf Lensing Signal}

To generate forecasts, we first need predictions for the expected amplitude of the lensing signal around dwarfs lenses. For this, we adopt the results from Leauthaud et al in prep. These are briefly summarized below.

The expected signal is generated directly from the Bolshoi Planck N-body simulation \citep{Klypin:2016ac}. The Bolshoi Planck simulation has a box size of Lbox=(250 Mpc/h)$^3$, 2048$^3$ particles, a particle mass of m$_p=1.5\times 10^8$ M$_{\odot}$, and resolves 10$^{10}$ M$_\odot$ halos. We use a snapshot at $a=0.78$ or $z=0.28$. A five parameter stellar-to-halo mass relation with mass dependent scatter was used to populate the Bolshoi simulation with mock galaxies down to $\log(M_*/M_{\odot})=8$. The parameters of this model were fit to the COSMOS stellar mass function (SMF) as well as to a galaxy-galaxy lensing signal measured for a dwarf sample with $\log(M_*/M_{\odot})\sim 8.5$. Both the SMF and the lensing help to ensure that the resulting mock catalog has a realistic population of dwarf galaxies. Further details are given in Leauthaud et al in prep.

Using the mock catalog described above, we can predict the amplitude of the galaxy-galaxy lensing observable ($\Delta\Sigma$) for dwarfs with $\log(M^*/M_{\odot})>8$. The predicted signal is computed from Bolshoi by selecting dwarfs in a narrow mass range and then cross-correlating this sample with the dark matter particles of the simulation. More specifically, we use the \textsf{delta\_sigma} function in \textsf{Halotools} \citep{Hearin:2017aa} to generate our model predictions.

We select mock galaxies in two narrow mass bins centered around $\log(M^*)=8$ and $\log(M^*)=9$. The predicted $\Delta\Sigma$ profiles are shown in Figure \ref{predicted}. This signal includes contributions from both central and satellite galaxies.

\subsection{Survey Parameters}

 Here we list the survey parameters assumed to generate forecasts. These numbers are also summarized in Table \ref{forecasttable}.
 
\begin{itemize}
\item For the HSC wide layer, we assume an area of 1000 deg$^2$, a source density of 18.5 galaxies per arcmin$^2$, and a mean source redshift of $z_s=0.81$ \citep{Hikage:2019aa}. 
\item For the main LSST survey, we assume an area of 18,000 deg$^2$, a source density of 30 galaxies per arcmin$^2$ with $z_s=1.2$ \citep{Chang:2013aa,Chan:2015aa}.
\item For the \emph{Euclid} wide layer we assume 15,000 deg$^2$, 30 galaxies/arcmin$^2$, and a mean source redshift of 0.9. 
\item Finally, we assume that the WFIRST High Latitude Survey (HLS) will observe 2,400 deg$^2$ and will yield a source density of 54 galaxies per arcmin$^2$ at a mean redshift of $z_s=1.1$. 
\end{itemize}
  
\subsection{Computation of Signal-to-noise}

We now use the survey parameters above to predict the errors on the $\Delta\Sigma$ profiles at $\log(M^*)=8$ and $\log(M^*)=9$. We assume one redshift bin extending from $z=0$ to $z=0.25$. The mean redshift of lenses is $\overline{z}=0.18$. We use the same COSMOS SMF as in Leauthaud et al in prep. to compute to number density of dwarfs within a given mass and redshift range.

Our methodology for computing the expected errors for $\Delta\Sigma$ follows \citet{Singh:2017}. We briefly summarize the salient features of this computation and refer the reader to \citet{Singh:2017} for further details. In short, the gaussian covariance for $\Delta\Sigma$ is given by:

\begin{displaymath}
  {\rm Cov}(\ds(r_p),\ds(r_p'))\approx  \frac{\Sigma_c^2(\chi_s,\chi_g)}{V_W} \int dk_\perp k_\perp J_2(k_\perp r_p)J_2(k_\perp r_p') \hspace{0.65\columnwidth}
\end{displaymath}
\begin{equation}
\label{cov}
 \left[(P_{gg}(k_\perp)+\frac{1}{n_g})(P_{\kappa\kappa}(k_\perp)+\frac{\sigma^2}{n_s})+P_{g\kappa}^2(k_\perp)\right],
\end{equation}

\noindent where $\chi_g$ and $\chi_s$ are the comoving distances to lens and source galaxies. We use the mean redshift for source galaxies as specified in Table \ref{forecasttable}. 

For the power spectrum, we use the \textsf{HaloFit} non linear power spectrum \citep{Smith:2003,Takahashi:2012aa}. For the galaxy power spectrum, we use linear galaxy bias with the non-linear matter power spectrum. The galaxy-matter power spectrum ($\Sigma_c^2P_{g\kappa}$) is obtained by direct inverse hankel transform of $\ds$.

The convergence power spectrum, $P_{\kappa\kappa}$ in units of $P(k)$ is given by:
\begin{equation}
    P_{\kappa\kappa}(k_\perp)=\int_0^{\chi_s}d\chi \frac{\overline{\rho}^2}{\Sigma_c^2(\chi_s,\chi)}P_{mm}(k_\perp\frac{\chi_g}{\chi})
\end{equation}

Our computation includes all terms relevant for the disconnected or gaussian covariance. However, we do not account for the effects of survey masks and selection functions, including the clustering of source galaxies. We also ignore contributions from the connected covariance which includes super sample covariance, as well as the trispectrum between galaxies and shear. We estimate that ignoring these teams will lead to S/N estimates that will be over optimistic by up to $\sim$25\%. To account for this, we apply a $\sim$25\% reduction in the S/N estimates reported in Table \ref{forecasttable2}.

\section{Results}\label{results}

Using the methodology described above, we compute the expected $\Delta\Sigma$ signal, and the errors on this signal. We consider two narrow mass bins centered at $\log(M^*)=8$ and $\log(M^*)=9$ and with a bin width of 0.2 dex. We assume one redshift bin from $0<z<0.25$ with lenses at a mean redshift of $\overline{z}=0.18$.  
 
\begin{figure*}
\centering
\includegraphics[width=16cm]{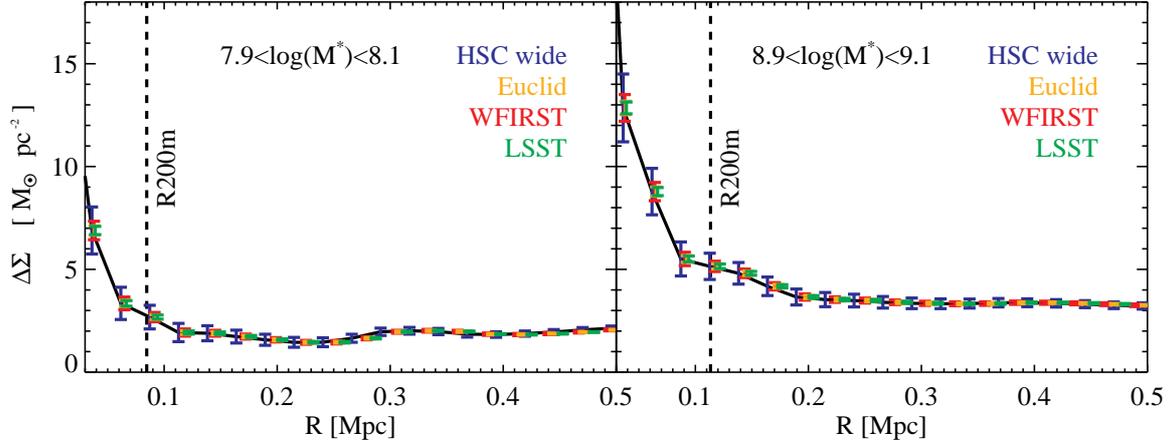}
\caption{Predicted lensing signal and errors for the $\Delta\Sigma$ profile around dwarf lens galaxies. The left hand panel corresponds to dwarfs with $\log(M^*)=8$ and the right hand panel corresponds to dwarfs with $\log(M^*)=9$. Errors correspond to a lens samples selected within a narrow mass range (0.2 dex bin width) and for $0<z<0.25$. Predicted diagonal errors are shown for HSC wide survey (blue), \emph{Euclid} (orange), WFIRST (red), and LSST (green).}
\label{predicted}
\end{figure*}

 Figure \ref{predicted} shows the expected amplitude of $\Delta\Sigma$ for these two mass bins and for radial scales below $R<500$ kpc. Figure \ref{predicted} also displays the predicted diagonal errors on $\Delta\Sigma$ for the HSC wide survey, \emph{Euclid}, WFIRST, and LSST. 

Based on our mocks, the mean halo mass of the $\log(M^*)=8$ sample is $\log(M_{\rm 200m})=10.91$ and the mean halo mass of the $\log(M^*)=9$ sample is $\log(M_{\rm 200m})=11.25$. The signal shown in Figure \ref{predicted} is therefore a combination of the dark matter halos associated with central dwarfs (the ``one halo" term at $R_{\rm 200m}<84.55$ kpc and $rR_{\rm 200m}=113.44$ kpc for $\log(M^*)=8$ and $\log(M^*)=9$ respectively), with satellite galaxies, and the signal arising from correlated structure at $r>R_{\rm 200m}$ (the so-called ``two halo term").

We now computed the expected S/N of the detections shown in Figure \ref{predicted} and report the corresponding values in Table \ref{forecasttable2}. We consider radial scales that correspond both to the one-halo term but also at $r<500$ kpc (physical) which also includes contributions from the two-halo term.   

The HSC wide survey will be the first with the capability of measuring the lensing signal for dwarfs with high signal-to-noise. The signal will be detected with enough significance to measure the signal in fine bins of mass (here the bins are only 0.2 dex in width). At $r<500$ kpc, the HSC wide survey will be able to measure the lensing signal with a signal-to-noise of 37 at $\log(M^*)=8$ and 46 at $\log(M^*)=9$. In the one-halo regime, the predicted S/N of the detection is 8 at $\log(M^*)=8$ and 15 at $\log(M^*)=9$.

Considering all of the lensing surveys taken together, LSST will have the most constraining power. We find that LSST will be able to measure the lensing signal with a signal-to-noise of 208 at $\log(M^*)=8$ and 261 at $\log(M^*)=9$ at $r<500$ kpc! In the one-halo regime, the predicted S/N of the detection for LSST is 47 at $\log(M^*)=8$ and 84 at $\log(M^*)=9$.

WFIRST and LSST will have the greatest capability of pushing
below the $\log(M^*)=7$ mass scale thanks to the depth of their imaging data. Exactly how low mass they will probe is likely to depend on surface brightness effects and whether or not the detection pipelines are optimized to detect faint and low surface brightness objects.

In order to disentangle baryonic effects from non baryonic solutions to the cusp-core controversy (such as self-interacting dark matter), it will become interesting to try and push the galaxy-galaxy lensing measurement down to the smallest radial scales possible to probe the inner dark matter profile. \citet{Kobayashi:2015} have shown that statistically speaking this ``small scale lensing" measurement is possible. However, pushing the lensing signal down to $r<$ 20 kpc will require the development of methods capable of  accurately measure the lensing signal in the presence of strong proximity and blending effects. On these very small scales ($R<100$ kpc), are forecasts are optimistic because they do not account for loss of source galaxies due of masking and blending effects.

Our results and the signal-to-noise values in Table \ref{forecasttable2} demonstrate that \emph{ that studies of the dark matter halos of dwarfs will not be limited by lensing signal-to-noise. Rather, lensing at the dwarf scale will be limited by our ability to accurately obtain redshifts for dwarf lenses}. 

We have shown in Section \ref{section:comp} that HSC, and future lensing surveys, will be deep enough to detect large samples of dwarfs. However, these surveys are photometric, and do not provide the redshifts that will be necessary to select low redshift dwarf lens samples. Further work will be required to study methods for obtaining redshifts. For example, it will be important to consider the feasibility of, and trade-offs between: wide field direct spectroscopic follow-up, prism/grism based redshifts, and narrow-band imaging follow-up \citep[e.g.][]{Eriksen:2019aa}. Finally, it will also be interesting to consider these in combination with machine learning methods for extracting dwarf samples from deep imaging surveys.

\begin{table*}
  \caption{Predicted signal-to-noise for two mass bins of width 0.2 dex and for $0<z<0.25$.}
\begin{tabular}{@{}lcccc}
\hline
Survey & $\log(M^*)$=8 and $r<500$ kpc & $\log(M^*)$=9 and $r<500$ kpc & $\log(M^*)$=8 and $r<84.55$ kpc & $\log(M^*)$=9 and $r<113.44$ kpc \\
\hline
HSC Wide & 37 & 46 & 8 & 15 \\
LSST Wide & 208 & 261 & 47 & 84\\
\emph{Euclid} Wide & 184 & 231 & 41 & 74 \\
WFIRST HLS & 92 & 153 & 21 & 37 \\
\hline
\end{tabular}
\label{forecasttable2}
\end{table*}

\section{Summary and Conclusions}\label{conclusions}

In this paper, we show that the advent of new photometric lensing surveys that are \emph{both deep and wide} will open up the possibility of direct measurements of the dark matter halos of dwarf galaxies with gravitational lensing. Deep photometry (i$\sim$26 mag) over wide areas ($A>1000$ deg$^2$) will enable the extraction of large enough samples of dwarf galaxies at $z\sim0.1$ to push galaxy-galaxy lensing measurements to the dwarf scale.

The HSC wide survey will be the first with the capability of measuring the lensing signal for dwarfs with high signal-to-noise. The signal will be detected with enough significance to measure the signal in fine bins of mass (here the bins are only 0.2 dex in width). LSST will have the most overall constraining power. We find that LSST will be able to measure the lensing signal with a signal-to-noise in excess of 200 at $\log(M^*)>8$. Finally, WFIRST and LSST will have the greatest potential for pushing below the $\log(M^*)=7$ mass scale.

HSC and other deep$+$wide lensing surveys will detect significant numbers of dwarf galaxies at $z\sim 0.1$. However, further work will be required in order to develop optimized strategies for determining redshifts and for extracting dwarfs samples from these surveys.

Studies of the dark matter halos of dwarf galaxies with gravitational lensing is soon within reach. The combination of small scale kinematics and weak lensing on larger scales will be a new powerful tool to constrain the halo masses and inner density slopes of dwarf galaxies and to distinguish between baryonic feedback and modified DM scenarios. 

\section*{Acknowledgements}

We thank Clotilde Laigle for providing the COSMOS point source sensitivity. We thank Henk Hoekstra for proving the \emph{Euclid} point source sensitivity. This material is based upon work supported by the National Science Foundation under Grant No. 1714610. This research was also supported in part by National Science Foundation under Grant No. NSF PHY11-25915. AL acknowledges support from the David and Lucille Packard foundation and from the Alfred P. Sloan foundation.


\bibliographystyle{mnras}
\bibliography{all_refs}
\label{lastpage}


\end{document}